\documentclass[twocolumn]{aastex62}

\usepackage{bm}

\graphicspath{{./}{figures/}}

\submitjournal{ApJ}

\shorttitle{Survival Time of Tight Planetary Systems}
\shortauthors{A. Yalinewich and C. Petrovich}

\begin{document}

\title{Nekhoroshev Estimates for the Survival Time of Tightly Packed Planetary Systems}

\correspondingauthor{Almog Yalinewich}
\email{almog.yalin@gmail.com}

\author{Almog Yalinewich}
\affil{Canadian Institute for Theoretical Astrophysics, 60 St. George St., Toronto, ON M5S 3H8, Canada}

\author{Cristobal Petrovich}
\affil{Canadian Institute for Theoretical Astrophysics, 60 St. George St., Toronto, ON M5S 3H8, Canada}
\affil{Centre for Planetary Sciences, Department of Physical \&  Environmental Sciences, University of Toronto at Scarborough, Toronto, Ontario M1C 1A4, Canada}
\affil{Steward Observatory, University of Arizona, 933 N. Cherry Ave., Tucson, AZ 85721, USA}


\begin{abstract}

$N$-body simulations of non-resonant tightly-packed planetary systems have found that their survival time (i.e. time to first close encounter) grows exponentially with their interplanetary spacing and planetary masses.
Although this result has important consequences for the assembly of planetary systems by giants collisions and their long-term evolution, this underlying exponential dependence is not understood from first principles, and previous attempts based on orbital diffusion have only yielded power-law scalings.
We propose a different picture, where large deviations of the system from its initial conditions is due to few slowly developing high-order resonances. Thus, we show that the survival time of the system  $T$ can be estimated using a heuristic motivated by Nekhoroshev's theorem, and obtain a  formula for systems away from overlapping two-body mean-motion resonances as: $T/P=c_1  \frac{a}{\Delta a} \exp \left(c_2 \frac{\Delta a}{a} /\mu^{1/4}\right)$, where $P$ is the average Keplerian period, $a$ is the average semi major axis, $\Delta a\ll a$ is the difference between the semi major axes of neighbouring planets, $\mu$ is the planet to star mass ratio, and $c_1$ and $c_2$ are dimensionless constants. We show that this formula is in good agreement with numerical N-body experiments for $c_1=5 \cdot 10^{-4}$ and $c_2=8$.

\end{abstract}



\section{Introduction} \label{sec:intro}

According to planet formation theory, small rocky cores are formed in the protoplanetary disk. These cores, as shown in computer simulation, are expected to be of similar mass and at a distance of a few Hill radii from one another \citep{Kokubo1998OligarchicProtoplanets}. After the disk evaporates, these rocky cores begin to collide and merge, eventually forming the terrestrial planets \citep{Agnor1999OnFormation, Chambers2001MakingPlanets}. The rate at which these collisions occur is of tremendous importance to planet formation and the survivability of planetary systems like our own. Collisions play a role, for example, in the Nice model \citep{Gomes2005OriginPlanets, Tsiganis2005OriginSystem, Morbidelli2005ChaoticSystem}, although there eccentricities are excited by the gas giants, whereas in this work we consider planetary systems with just terrestrial planets.

\begin{figure*}
\begin{center}
	\includegraphics[width=\textwidth]{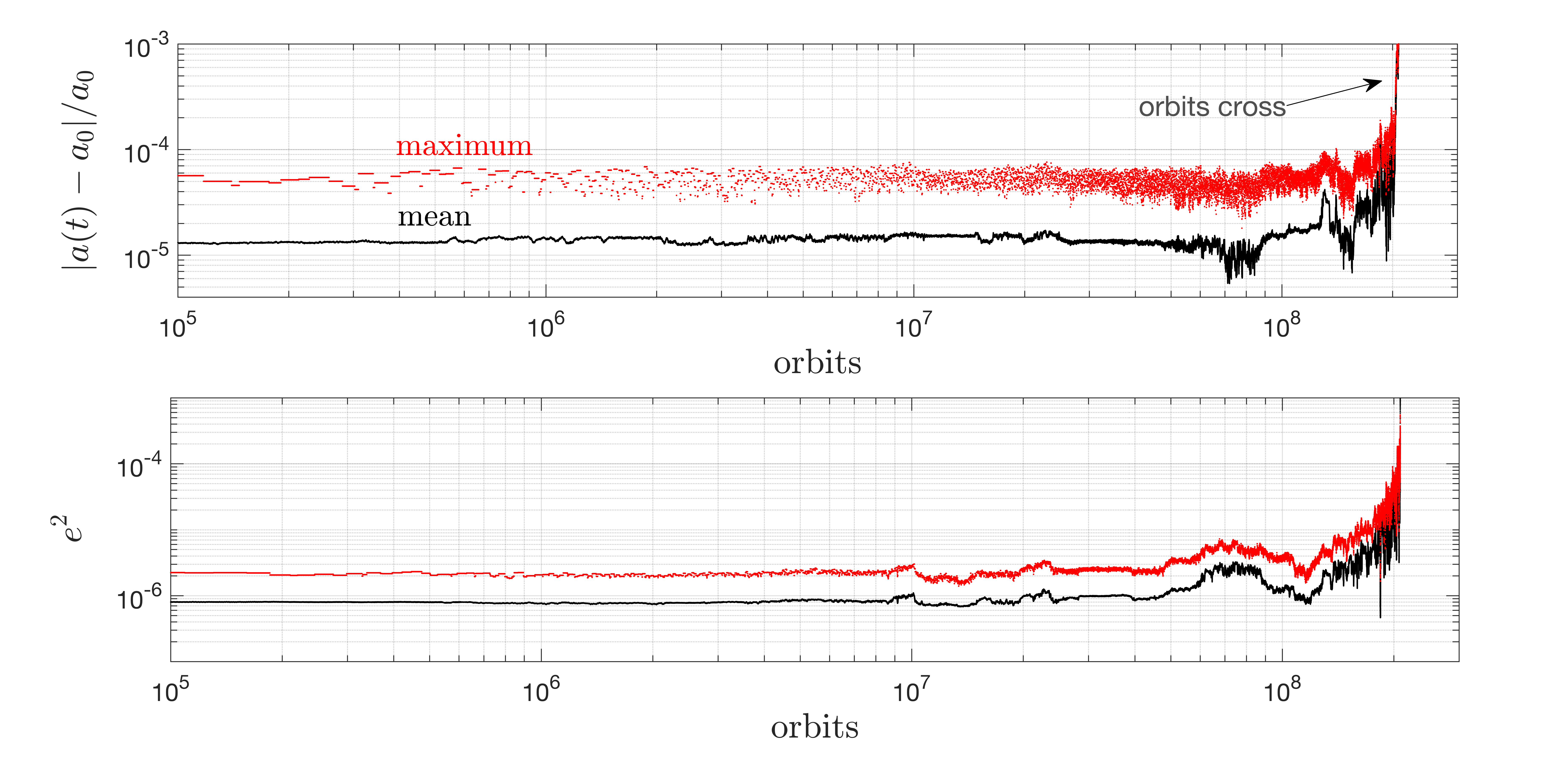}
    \caption{Time evolution of the relative changes in semi-major axes and eccentricity squared for the third planet in a four planet system that becomes unstable after $\sim 2\times10^{8}$ orbits. These variables are proportional to the Poincar\'e's actions $\left|\Lambda(t) - \Lambda_0 \right| \simeq \Lambda_0 |a(t)- a_0|/(2 a_0)$ and $\left|\Gamma (t) - \Gamma_0\right| \simeq \Lambda_0 e^2/2$.
    The dots corresponds to the mean and maximum values using a time resolution of 100 orbits and a moving window of $10^{4}$ orbits to smooth-out the high frequency oscillations.
    We observe that for the majority of time the action variables are bounded, until a sharp rise occurs. In contrast, if the system were to evolve by diffusion, we would expect to see a steady, power-law increase of the action variables with time. In this examples set the mass ratio between the planets and the star to be $\mu=10^{-7}$, equal spacing of $(a_{i+1}-a_{\rm i})/a_1=0.046$ (nearly 10 mutual Hill radii) and eccentricities to $e_i=10^{-7}$, chosen such that no pair of first-order resonances overlap according to the widths calculated by \cite{Deck2012RAPIDSYSTEM}. The true anomalies and longitudes of periapse are set to $\{0,-1.4,1,-1.4\}$ and $\{0.9,3.7,3.3,3.7\}$ for the planets \{1,2,3,4\}.
    The integration is carried out using the WHfast package \citep{Rein2015WHFast:Simulations} in Rebound \citep{Rein2012REBOUND:Dynamics} with $\simeq 50$ timesteps per orbit.}
    \label{fig:action_variable_evolution}
\end{center}
\end{figure*}

The stability of planetary systems in general, and of our solar system in particular, is a long standing problem in physics and mathematics \citep{Laskar1989ASystem, Sussman1992ChaoticSystem, Murray1999TheSystem}, which has been studied extensively, both analytically \citep{Celletti2007KAMMechanics, Efthymiopoulos2008OnDiffusion, FEJOZ2004DemonstrationHerman, Giorgilli2009KolmogorovBodies, Giorgilli2017SecularTheories, Locatelli2007InvariantSystem, Robutel1995StabilityProblem, Sansottera2010OnSystem} and numerically \citep{Batygin2008OnSystem, Hayes2007IsChaotic, Laskar1994Large-scaleSystem., Laskar2009ExistenceEarth}. The chaotic evolution of such systems makes their dynamics difficult to analyze. To illustrate this issue, we have plotted the time evolution of the eccentricity and semi-major axis of one planet in a tightly packed four planet system in Figure \ref{fig:action_variable_evolution}. The planets end up colliding, but rather than increasing gradually, the eccentricity and semi-major axis seem to be bounded for about $10^8$ orbits, and then shoot up at a seemingly arbitrary time.

The major analytic progress in this field is the celebrated Kolmogorov-Arnol'd-Moser (KAM) theory \citep{Kolmogorov1954OnFunction, Arnold1963PROOFHAMILTONIAN, Moser1962OnAnnulus, Poschel2009ATheorem}. The main result of this theory is that for small enough perturbations to non degenerate, non resonant integrable Hamiltonian system, the motion lies on an invariant torus with fixed frequencies. One of the supplements to the KAM theory is the Nekhoroshev theory \citep{Nekhoroshev1977ANSYSTEMS, Nekhoroshev1979AnSystems, Poschel1993NekhoroshevSystems, Niederman2012NekhoroshevTheory}, which states that even if a system does not satisfy the condition for KAM stability, the deviations from the unperturbed solution remain bounded for exponentially long times.

In its original formulation, Nekhoroshev's estimates apply to non degenerate systems, whereas the planetary problem is degenerate. Various authors have shown that a canonical change of coordinates can circumvent the degeneracy \citep{Niederman1996StabilityProblem, Guzzo2007AnTheorem}, and applied the Nekhoroshev theory to study the long-term orbital stability of bodies in our Solar system \citep{Cellett1996AnProblem, Morbidelli1997TheSystem}.  \cite{Pavlovic2008FulfillmentFamilies} also showed that the Nekhoroshev theorem is applicable to the Koronis and Veritas asteroid families.
We use a similar method to study long-term evolution of planetary systems, but focused on compact extra-solar systems for which the relative separation is comparable to their Hill radius. Such compact configurations are expected for rocky cores soon after the protoplanetary disk evaporates, while some systems might retain this compactness for billions of years as it has been revealed by the Kepler sample \citep{Pu2015SPACINGINSTABILITY}.

The stability of tight multi-planet systems with more than two planets has been explored primarily using numerical N-body simulations \citep[e.g.][]{Chambers1996TheSystems, Yoshinaga1999TheSystems, Zhou2007Post-OligarchicSystems, Chatterjee2014PlanetesimalPlanets, Smith2009OrbitalPlanets, Funk2010TheSystems}.  These simulations reveal that the survival time of the system---the time until the first collision happens---increases very steeply (exponentially) with the initial separation between the orbits \citep[see][for a summary of previous numerical  results]{Pu2015SPACINGINSTABILITY}. This behavior, which has been reproduced in multiple independent studies, is not understood theoretically \citep[see attempts by][]{Chambers1996TheSystems, Zhou2007Post-OligarchicSystems, Quillen2011Three-bodySystems}.

In this paper we propose that this survival time can be estimated using Nekhoroshev's theory. For simplicity, we make a number of assumptions. We assume that all the planets are of equal mass move in co-planar, circular, prograde (i.e. all planets orbit in the same direction), regularly spaced orbits. We also assume that the system does not start out in any first order two body resonance \citep{Deck2012RAPIDSYSTEM}.

The plan of the paper is as follows: in section \ref{sec:math} we present Nekhoroshev's theory and use it to obtain a survival time for tight planetary systems. In section \ref{sec:sim} we compare our results to numerical N-body simulations. Finally, in section \ref{sec:discussion} we discuss our results.

\section{Mathematical Model} \label{sec:math}

\subsection{Nekhoroshev Estimates} \label{sec:nekhoroshev_estimates}

We consider a nearly integrable Hamiltonian system of the form
\begin{equation}
    H \left(\bm{I}, \bm{\varphi}\right) = H_0 \left(\bm{I}\right) + \varepsilon H_1 \left(\bm{I}, \bm{\varphi}\right) \, \label{eq:perturbed_hamiltonian}
\end{equation}
with $N$ degrees of freedom, $\varepsilon \ll 1$ measures the strength of the perturbation, and $\bm{I}$ and $\bm{\varphi}$ are action-angle variables. We note that when $\varepsilon = 0$ then all $I$ are constant and all angles increase linearly with time. If $H_0 \left(\bm{I}\right)$ satisfies certain reasonable conditions on its convexity, and if $\varepsilon$ is small enough, then it can be shown that the system remains stable for exponentially long times. Mathematically, it is stated in the following way: the deviation from the unperturbed solution is bounded from above by some power law of the perturbation strength
\begin{equation}
    |\bm{I}\left(t\right) - \bm{I}\left(0\right)| < C_1 \varepsilon^{\frac{1}{2N}}
\end{equation}
for a duration of time that increases exponentially as the strength of the perturbation decreases
\begin{equation}
    t < C_3 \exp \left[C_2 / \varepsilon^{\frac{1}{2N}}\right]
\end{equation}
where $C_1$ and $C_2$ are constants that do not depend on $\varepsilon$ or time, and $C_3$ can, at most, be polynomial in $\varepsilon$. This constraint is known as the Nekhoroshev estimate. We emphasise that this condition guarantees that the system remains bounded within a certain time period, but afterwards the system can either remain stable or deviate further from the unperturbed solution. 
We note that since these are upper bounds, different authors arrive at different power law indices of $\varepsilon$, depending on their assumptions about the behavior of the Hamiltonian \citep{Guzzo2007AnTheorem}.

A complete proof of this constraint is given in \cite{Lochak1992EstimatesHamiltonian}, and we will not repeat it here. Instead, we only review some of the key ideas behind the proof. First, we express the perturbing function as a Fourier transform in the angle variables
\begin{equation}
    H_1 \left(\bm{I}, \bm{\varphi}\right) = \sum_{k} h_{\bm{k}} \left(\bm{I}\right) \exp \left(i \bm{k} \cdot \bm{\varphi}\right).
\end{equation}
where $\bm{k}$ represents a list of integers. Expanding Hamilton's equation for the momenta to first order in the small parameter $\varepsilon$ we obtain
\begin{equation}
    \dot{\bm{I}}_1 = - \nabla_{\bm{\varphi}}H \approx i \varepsilon \sum_{\bm{k}} h_{\bm{k}} \left(\bm{I}_0 \right) \bm{k} \exp \left(i \bm{k} \cdot \bm{\varphi}_0\right)
\end{equation}
where $\bm{I}_0$ are constants and $\bm{\varphi}_0 = \bm{\omega} t + \bm{\varphi}_i$, $\bm{\varphi}_i$ are constants, and $\bm{\omega} = \nabla_{\bm{I}} H_0 |_{\bm{I}=\bm{I}_0}$. Time integration yields
\begin{equation}
    \bm{I}_1 = \varepsilon \sum_{\bm{k}} h_{\bm{k}} \left(I_0\right) \bm{k} \frac{\exp \left(i \bm{k} \cdot \bm{\varphi}_0\right)}{\bm{k} \cdot \bm{\omega}}. \label{eq:first_order_action}
\end{equation}
We immediately notice a problem with this approach. For arbitrarily large values of the integers $\bm{k}$, the product $\bm{k} \cdot \bm{\omega}$ can be brought arbitrarily close to zero. In fact, a number theory  result called Dirichlet's approximation theorem states that the convergence is better than $|\bm{k}|^{1-n}$, where $n$ is the number of integers and $|\bm{k}| = \sum_l |k_l|$ is the sum of their magnitudes. Terms for which $\bm{k} \cdot \bm{\omega} = 0$ are called resonant terms. Non-resonant terms oscillate in time, while resonant terms grow linearly with time.

On the other hand, the Paley-Wiener theorem tells us that at large indices $\bm{k}$, the Fourier coefficients must decline at least exponentially. Therefore, resonant terms eventually grow and invalidate the perturbative solution, but they take at least an exponentially long time to grow. The remaining challenge is to estimate the index limit $|\bm{k}|$.

\subsection{Adaptation to Tightly Packed Planetary Systems}

The physical system we are interested in is a small number of planets of the same mass $m$ around a star of mass $M \gg m$. The average distance between the star and the planet is $a$ and the initial separation between neighbouring planets' semi major axes is $\Delta a \ll a$. All planets are assumed to initially move on circular, coplanar orbits (i.e. both positions and velocities are coplanar, so the motion is restricted to a single plane). We want to estimate the average time it takes for the first planet-planet collision to occur, which it translates into a condition on the growth of the action $I$ (either semimajor axes or eccentricities). 

In order to illustrate the evolution of such systems and we show an example in Figure 1 for a four planet system. This example shows that the evolution the actions remain bounded for a long time and after $\sim 10^{8}$ orbits, they break loose and diffuse to the absorbing boundaries (i.e., orbit crossing).

Hereafter, we will use Poincar\'e variables, the action-angle pairs for a given planet are $\{\Lambda\simeq m\sqrt{GMa},\lambda\}$ and $\{\Gamma\simeq m\sqrt{GMa} e^2/2, -\varpi \}$ with $\lambda$ and $\varpi$ the mean longitude and the longitude of pericenter, respectively \citep{Murray1999SolarDynamics}.

\subsubsection{Scaling with Spacing $\Delta a/a$}
From the discussion in the previous section, we know that this timescale will depend on the exponential decline of the Fourier coefficients. Indeed, the gravitational binding energy between two planets is given by
\begin{equation}
    H_1 \propto \frac{1}{\sqrt{r_1^2 + r_2^2 - 2 r_1 r_2 \cos \theta}} \propto \sum_j b_{\frac{1}{2}}^j \left(\frac{r_2}{r_1}\right) \cos \left(j \theta\right)
\end{equation}
where $r_{1,2}$ are distances of the planets from the host star, $\theta$ is the angle between the lines connecting the planets to the star, and
\begin{equation}
    b_{\frac{1}{2}}^j \left(\alpha\right) = \frac{1}{\pi} \int_0^{2 \pi} \frac{\cos \left(j \theta\right)d\theta}{\sqrt{1 - 2 \alpha \cos \theta + \alpha^2}}
\end{equation}
are the Laplace coefficients, which decline exponentially with the index  $b_{\frac{1}{2}}^j \left(\alpha\right) \propto \exp \left(-\left(\alpha-1\right) |j|\right)$ \citep{Quillen2011Three-bodySystems}. Thus, for nearly circular orbits:
\begin{equation}
b_{\frac{1}{2}}^j \left(\alpha\right) \propto \frac{1}{2} \left|\ln \frac{\Delta a}{a} \right| \exp \left[-\frac{\Delta a}{a} |j|\right]
\end{equation}
(though we will neglect the logarithmic term, as it only introduces a small correction) and from Equation (\ref{eq:first_order_action}) we know that the deviation from the unperturbed solution in the resonant case to grow linearly with time roughly as 
\begin{equation}
|\bm{I}_1| \propto t \cdot h_{\bm{j}} \left(\bm{I}_0\right) \propto t \cdot \exp \left[-\frac{\Delta a}{a} |j|\right] \label{eq:I_1_growth}
\end{equation}
The system survives as long as the action variable does not exceed some critical threshold $|\bm{I}_1| < I_{\rm crit.}$, and from Equation (\ref{eq:I_1_growth}) the survival time $T$ scales as 
\begin{equation}
\ln T \propto \frac{\Delta a}{a}.
\end{equation}

\subsubsection{Scaling with the Mass Ratio $\mu=m/M$}

We now turn to calculate the dependence of the survival time on the mass ratio. Unlike the dependence with the spacing, this requires specifying the nature of the perturbations that lead to growth of the actions.

We shall start by assuming that the evolution of a single planetary orbit is mainly determined by its two closest neighbors (thus neglecting resonances with non adjacent bodies) so the relevant perturbative Hamiltonian involves the gravitational interactions of only three planets. This choice is justified by the fact that the strength of interactions, which is proportional to the Laplace coefficients and decays exponentially with interplanetary separation \citep{Quillen2011Three-bodySystems}.

Away from overlapping two-body resonances,  \citet{Quillen2011Three-bodySystems}  showed that three-body resonances dominate the chaotic dynamics of low-mass multi-planet system. To first-order in the eccentricities, this Hamiltonian can be written as
\begin{equation}
    H \left(\bm{\Lambda},\bm{\Gamma}, \bm{\lambda},\bm{\varpi}\right) = H_0 \left(\bm{\Lambda},\bm{\Gamma}\right) + \varepsilon H_1 \left(\bm{\Lambda},\bm{\Gamma}, \bm{\lambda},\bm{\varpi}\right) \, \label{eq:perturbed_hamiltonian_2}
\end{equation}
where $\bm{ \Lambda}=(\Lambda_1,\Lambda_2,\Lambda_3)$, $\bm{ \Gamma}=(\Gamma_1,\Gamma_2,\Gamma_3)$, $\bm{ \lambda}=(\lambda_1,\lambda_2,\lambda_3)$, 
$\bm{ \varpi}=(\varpi_1,\varpi_2,\varpi_3)$, and the small parameter is  $\varepsilon = \left(m/M\right)^2$, since three body resonances do not occur in the linear expansion in the mass ratio. After canonical transformations, this Hamiltonian has four degrees of freedom (since two were eliminated by the introduction of resonant angles) and can be expressed as 
\begin{equation}
    H \left(\bm{\Theta}, \bm{\theta}\right) = H_0 \left(\bm{\Theta}\right) + \varepsilon H_1 \left(\bm{\Theta}, \bm{\theta}\right) \, \label{eq:perturbed_hamiltonian_3}
\end{equation}
where $\bm{\Theta}$ is a function of the Poincar\'e actions and $\bm{\theta}$ a linear combination of $\bm{\lambda}$ and $\bm{\varpi}$. In particular, including the modes that are zeroth- and first-order in the eccentricities, one possible combination is \citep{Quillen2011Three-bodySystems}
\begin{eqnarray}
\theta_1&=&(p+1)\lambda_1-(p+q)\lambda_2+q\lambda_3-\varpi_1 \nonumber,\\
\theta_2&=&(p+1)\lambda_1-(p+q)\lambda_2+q\lambda_3-\varpi_2 \nonumber,\\
\theta_3&=&p\lambda_1-(p+q-1)\lambda_2+q\lambda_3-\varpi_2 \nonumber,\\
\theta_4&=&p\lambda_1-(p+q-1)\lambda_2+q\lambda_3-\varpi_3. 
\end{eqnarray}
Thus, since this Hamiltonian involves $N=4$ degrees of freedom,  Nekhoroshev's theorem predicts the survival time scales as
\begin{equation}
\ln T \propto \varepsilon^{-1/(2N)} \propto \left(\frac{M}{m}\right)^{1/4}.
\end{equation}

Our argument to get scaling with the mass assumes three-body interactions. Alternatively, there might be other set of perturbations involving two-body interactions that can destabilize the system. Since two-body interactions are linear in the masses, $\epsilon=m/M$, a two-degree-of-freedom model would reproduce the same scaling.  We remark, however, that our analysis excludes overlapping two-body resonances and these, as clearly shown in simulations by \citet{Obertas2017TheStar} with uniform spacing\footnote{Uniform spacing leads to nearly uniform period ratio distribution enhancing the appearance of overlapping two-body resonances.}, produce wild oscillations in the survival times. In turn, we suspect that the weaker, but much denser (abundant set of angles above that can resonate), three-body resonances would map into a very smooth distribution of $T$ with mutual spacing (high density in $\Delta a/a$) as observed in simulations with unequal spacing \citep[e.g.][]{Chambers1996TheSystems, Faber2007TheClearings}.

In summary, our estimates using three-body interactions and the Nekhoroshev's theorem results in a scaling with the mass as $\ln T  \propto \left(M/m\right)^{1/4}$. This scaling differs from the commonly used Hill scaling (i.e., $\ln T  \propto \left(M/m\right)^{1/3}$). We note, however, that previous simulations reported better fits with $\left(M/m\right)^{1/4}$ than that of the Hill's scaling \citep[e.g., Figure 4 of][]{Chambers1996TheSystems}. A more in-depth study on the mass scaling can better test our results.

\subsection{The Prefactor}

The remaining component for the survival time is the prefactor which contains a time scale. In this context, the relevant time scale is the time between conjunctions of neighboring planets, or the synodic time. Both planets orbit the star at roughly the Keplerian time $t_k \approx \sqrt{\frac{a^3}{G M}}$, and the relative difference in periods is proportional to $\Delta a/a$, so the synodic time is given by $t_s \approx \frac{a}{\Delta a} t_k \approx \frac{a}{\Delta a} \sqrt{\frac{a^3}{G M}}$. 

Finally we are able to write a complete expression for the survival time of tight planetary systems
\begin{equation}
    T = c_1 \frac{a}{\Delta a} \sqrt{\frac{a^3}{G M}} \exp \left(c_2 \frac{\Delta a}{a} \left(\frac{M}{m}\right)^{1/4} \right) \label{eq:survival_time}
\end{equation}
where, again, $c_1$ and $c_2$ are numerical constants that depend neither on $\Delta a/a$ nor $m/M$, and cannot be determined from scaling arguments. In the next section we compare this formula to numerical results, and using those results we are able to calibrate the numerical coefficients $c_1$ and $c_2$.

\section{Comparison to Simulations} \label{sec:sim}

In this section we compare our theoretical prediction in Equation (\ref{eq:survival_time}) to numerical N-body simulations of tight planetary systems. The results of most simulations done in the past are summarized in \cite{Pu2015SPACINGINSTABILITY}, but here we simply highlight specific works that explore relevant parts for our discussion.

The first set of simulations we consider was performed in \citep{Zhou2007Post-OligarchicSystems}. They simulated nine planets, spaced with a constant Hill parameter $K \approx 2 \frac{a_{i+1}-a_{i}}{a_{i+1}+a_i} \left(\frac{M}{m}\right)^{1/3}$. They found a very steep dependence of the survival time on the Hill parameter. They fit this dependence to a power law in the Hill parameter, and infer $d \ln T / d \ln K \approx 20$. They also varied the mass ratio $\mu = m/M$, and find $d \ln T/d \ln \mu = -0.27$. They developed an analytic theory for diffusion in eccentricity that reproduces the dependence on the mass ratio $\mu$, but whose dependence on the Hill parameter is too shallow ($K^5$ instead of $K^{20}$). A discussion of this theory and its shortcomings is given in the appendix.

A similar suite of simulations was performed by \cite{Rice2018SurvivalEncounter}. They explored a wider range of Hill parameters than \citet{Zhou2007Post-OligarchicSystems}, and they fit the numerical result to an exponential relation between the survival time and Hill parameter. A similar fit to an exponential was obtained by \citet{Petrovich2015TheSystems} for the case of two eccentric planets. From the results of \citet{Rice2018SurvivalEncounter} we are able to constrain the remaining free parameters $c_1$ and $c_2$ to obtain a final form for the survival time of tight planetary systems
\begin{equation}
    T/P \approx 5 \times 10^{-4} \cdot \left( \frac{a}{\Delta a}\right) \exp \left(8 \left(\frac{\Delta a}{a}\right) \left(\frac{M}{m}\right)^{1/4}\right) \,, \label{eq:final}
\end{equation}
or more commonly expressed as
\begin{equation}
    \log(T/P) \approx -4.4+ \log\left(\frac{a}{\Delta a}\right) +3.5\; \left(\frac{\Delta a}{a}\right) \left(\frac{M}{m}\right)^{1/4}. \label{eq:final2}
\end{equation}
This form is similar to the expression found by \citet{Faber2007TheClearings}, which has a coefficient of 3.7 accompanying the term $\frac{\Delta a}{a} \left(\frac{M}{m}\right)^{1/4}$ instead of 3.5, and logarithmic dependence on the mass ratio, not the spacing.

\begin{figure}
	\includegraphics[width=0.5\textwidth]{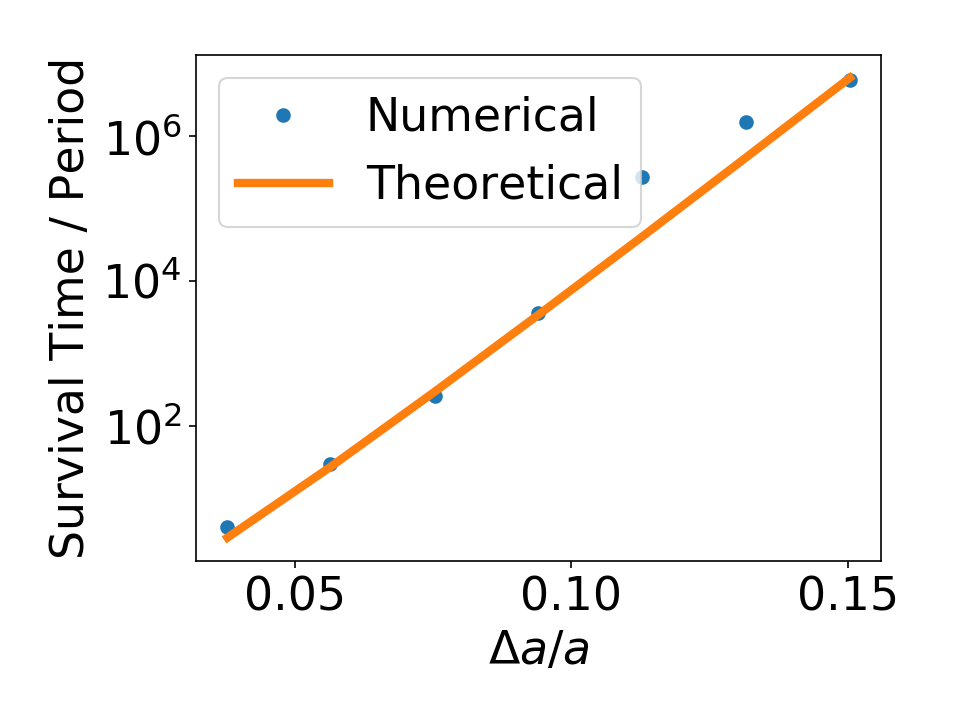}
    \caption{A comparison between the numerical results of \citet{Rice2018SurvivalEncounter} for the relation between the survival time and the interplanetary spacing and a fit to these results using our analytical functional form in Equation (\ref{eq:survival_time}) (i.e.,
determining the two constants $c_1$ and $c_2$). The best fit results  in Equation (\ref{eq:final}) and agrees fairly well with the numerical experiments.}
    \label{fig:sim_comparison}
\end{figure}

A comparison between the numerical results of \citet{Rice2018SurvivalEncounter} and Equation (\ref{eq:final}) can be seen in figure \ref{fig:sim_comparison}. We note that \citet{Rice2018SurvivalEncounter} argue that the argument in the exponent should be proportional to the Hill parameter $\frac{\Delta a}{a} \left(\frac{M}{m}\right)^{1/3}$, unlike our results. However, we note that some numerical experiments agree well with the $\frac{\Delta a}{a}\left(\frac{M}{m}\right)^{1/4}$ scaling \citep{Chambers2001MakingPlanets,Faber2007TheClearings}. Similarly, this scaling is born out from theoretical works estimating the regions for the onset of chaos either from three-body resonance overlap \citep{Quillen2011Three-bodySystems} or two-body resonance overlap for eccentric planets \citep{Hadden2018APlanets}. A more exhaustive study on the mass scaling of instability times will shed light on this issues.

Finally, \citet{Obertas2017TheStar} ran another suite of simulations, with a much higher resolution in the interplanetary spacing compared to all previous works. They found a general trend consistent with the previous studies, and also large and rapid variations in the survival time as a function of $\Delta a/a$. They show that the locations of these oscillations coincide with two-body mean-motion resonances. From the discussion in section \ref{sec:nekhoroshev_estimates} we can understand why our model fails for systems close to a resonance. In such cases, the Fourier series for the perturbing potential must be truncated at a indices smaller than $|\bm{k}| \propto \varepsilon^{-1/(2N)}$. We note that a resonance is a double edged sword, in the sense that it can rapidly change the parameters of the system, but only within a lower dimensional subspace of parameter space. If a collision is possible within this subspace, then it can happen on timescales much shorter than Nekhorosev's estimates. On the other hand, if a collision is not possible within this parameter space, then a resonance can prolong the lifetime of the system considerably. Both of these effects can be seen in the results of \citet{Obertas2017TheStar}.

\section{Discussion} \label{sec:discussion}

In this work we study the survival time (time to first close encounter) of tight planetary systems (mutual spacing $\lesssim 10$ mutual Hill radii).
We use Nekhoroshev's stability limit to derive an analytic formula for the  survival time as a function of interplanetary spacing and planet-to-star mass ratio. This formula is defined up to two dimensionless numbers and it reproduce the exponential scaling with interplanetary spacing and mass ratio observed several numerical simulations. By calibrating these constants we arrive at a complete expression for the survival time of tight planetary systems (Equation \ref{eq:final}).

We note that most previous attempts to explain the steep dependence of the survival time on the separation focused on estimating the chaotic diffusion coefficients, which only led to polynomial growth of the survival time. To our knowledge, this is the first work attempting to explain the numerical results with an exponential, although the idea has been suggested in previous works \citep{Zhou2007Post-OligarchicSystems, Quillen2011Three-bodySystems}. In our picture, a high order resonant mode grows linearly with time, but has a tiny prefactor, so it takes a long time for it to exceed low order non resonant modes.

We note that while this work provides a useful tool for estimating the survival time for tight planetary systems, our arguments are heuristic and do not reveal the nature of the chaotic interactions leading to instability. We suggest that three-body resonances are suspect, but have not explored the role of two-body resonances. Based on recent numerical simulations showing large deviations from our results in the vicinity of overlapping mean-motion resonances, and it is also possible that these resonant systems follow a different scaling law, possibly linked to rapid diffusion.
Finally, in this work we have assume a highly idealized system in which all planets have equal masses and uniformly spaced in coplanar orbits (i.e. both positions and velocities are coplanar, so the motion is restricted to a single plane). All these effects merit further exploration.

\section*{Acknowledgments}
We would like to thank Norm Murray and Alysa Obertas for the useful feedback. AY is supported by the Vincent and Beatrice Tremaine Fellowship. C.P. acknowledges support from the Gruber Foundation Fellowship and Jeffrey L. Bishop Fellowship. This work made use of the matplotlib python package \citep{Hunter2007Matplotlib:Environment}.

\bibliographystyle{apalike}
\bibliography{references}

\appendix

\section{Diffusion timescale of the eccentricity vector: understanding the power-law scaling using the impulse approximation}
In this section we discuss the analytic theory for diffusion in eccentricity developed in \citep{Zhou2007Post-OligarchicSystems}. The original derivation relies on a Hamiltonian formalism, but we can reproduce the results using the impulse approximation. In this approximation, we neglect all interaction between the planets until conjunction, and then assume all exchange of energy and momentum is instantaneous. Again, we assume initially co-planar planets on circular orbits. The mass of the planets is $m$, the mass of the star is $M$, the average semi major axis is $a$, and the separation between the planets is $\Delta a \ll a$.

Both planets move at roughly the Keplerian velocity $v_k \approx \sqrt{G M/a}$, and the relative velocity between them is $\Delta v_k \approx \frac{\Delta a}{a} \sqrt{\frac{G M}{a}}$. At conjunction, the acceleration between the two planets due to mutual attraction is $G m/\Delta a^2$. The duration of the interaction is roughly given by the time during which the distance between the two planets is $\Delta a$, namely $\Delta t \approx \Delta a/\Delta v_k \approx t_k \approx \sqrt{\frac{a^3}{G M}}$. Hence, due to the interaction, the planets receive a kick velocity normal to the direction of velocity difference
\begin{equation}
    \delta v_{\perp} \approx \frac{G m}{\Delta a^2} \sqrt{\frac{a^3}{G M}} \label{eq:radial_kick}
\end{equation}
In the case of a perfectly circular orbit, this kick velocity does not change the angular momentum, since the velocity is parallel to the radial direction. If, however, the orbit has an eccentricity $e$, and conjunction happens at some random angle, then the angle between the kick velocity and the radial direction would be proportional to the eccentricity $e$, and the kick velocity would have a small component in the tangential direction
\begin{equation}
    \delta v_t \approx e \delta v_{\perp} \approx e \frac{G m}{\Delta a^2} \sqrt{\frac{a^3}{G M}}
\end{equation}
Depending on the true anomalies at conjunction, this kick could either be prograge or retrograde. The change in angular momentum is given by $\Delta L \approx a \delta v_t$, and the change in eccentricity is
\begin{equation}
    \delta e^2 \approx \frac{\delta L}{L} \approx e \delta v_t/v_k \Rightarrow \delta e \approx \frac{m}{M} \frac{a^2}{\Delta a^2}
\end{equation}
where $L \approx \sqrt{G M a}$ is the orbital angular momentum per unit mass. For the orbits to cross, the eccentricity has to grow to $e_f \approx \Delta a/a$. Since we assume a random walk, the number of conjunction needed is $\left(e_f/\delta e\right)^2$. Since the time between conjunctions is $t_s \approx t_k a/\Delta a$, the survival time according to this model is
\begin{equation}
    T \approx t_s \left(\frac{e_f}{\delta e}\right)^2 \approx \sqrt{\frac{a^3}{G M}} \left(\frac{\Delta a}{a}\right)^5 \left(\frac{M}{m}\right)^2.
\end{equation}
Thus we reproduce the expression for survival time obtained by \citet{Zhou2007Post-OligarchicSystems}. This expression does not reproduce the observed exponential dependence of $T$ with  the spacing and the mass ratio. 

The problem with this model is that it assume a random angle between planets at conjunction, whereas the planets quickly align their elliptical orbits. To see why this is the case, let us use the impulse approximation to calculate how the direction of periapse changes. We already calculate the kick velocity in the radial direction in equation \ref{eq:radial_kick}. This component is always in the same direction, and does not depend on the eccentricity. We recall that the eccentricity vector is given by $\vec{e} = \frac{\vec{v}\times \vec{L}}{GM} - \hat{r}$, where $\vec{v}$ is the orbital velocity and $\vec{L}$ is the angular momentum. The change in this vector is given by 
\begin{equation}
    \delta \vec{e} = \pm \frac{m}{M} \frac{a^2}{\Delta a^2} \hat{\theta}
\end{equation}
where $\hat{\theta}$ represents the tangential direction. In this case the length of the eccentricity vector does not change, only its direction. We note that since the inferior and superior planets get opposite radial kick velocities but have roughly the same angular momentum vector, then the changes in the eccentricity vectors have opposite signs. If both planets had the same eccentricity vector, this would cause them to rotate in opposite ways. On the other hand, if both planets had equal but opposite eccentricity vectors, then this would cause both to rotate in the same way, thus preserving their anti-alignment. This anti-alignment can be maintained even for a large mass ratio between the planets, since in this case the more massive planet would have a much smaller eccentricity, which can be easily rotated by the smaller planet.

\end{document}